\input  eplain
\beginpackages                 
\usepackage{graphicx}
\usepackage{color}
\endpackages
\enablehyperlinks
\hlopts{bwidth=0,colormodel=,color=blue}


\catcode`@=11 


\font\ninerm=cmr9
\font\eightrm=cmr8
\font\sixrm=cmr6

\font\ninei=cmmi9
\font\eighti=cmmi8
\font\sixi=cmmi6
\skewchar\ninei='177 \skewchar\eighti='177 \skewchar\sixi='177

\font\ninesy=cmsy9
\font\eightsy=cmsy8
\font\sixsy=cmsy6
\skewchar\ninesy='60 \skewchar\eightsy='60 \skewchar\sixsy='60

\font\ninebf=cmbx9
\font\eightbf=cmbx8
\font\sixbf=cmbx6

\font\ninett=cmtt9
\font\eighttt=cmtt8

\hyphenchar\tentt=-1 
\hyphenchar\ninett=-1
\hyphenchar\eighttt=-1

\font\ninesl=cmsl9
\font\eightsl=cmsl8

\font\nineit=cmti9
\font\eightit=cmti8



\newskip\ttglue
\def\tenpoint{\def\rm{\fam0\tenrm}%
  \textfont0=\tenrm \scriptfont0=\sevenrm \scriptscriptfont0=\fiverm
  \textfont1=\teni \scriptfont1=\seveni \scriptscriptfont1=\fivei
  \textfont2=\tensy \scriptfont2=\sevensy \scriptscriptfont2=\fivesy
  \textfont3=\tenex \scriptfont3=\tenex \scriptscriptfont3=\tenex
  \def\it{\fam\itfam\tenit}%
  \textfont\itfam=\tenit
  \def\sl{\fam\slfam\tensl}%
  \textfont\slfam=\tensl
  \def\bf{\fam\bffam\tenbf}%
  \textfont\bffam=\tenbf \scriptfont\bffam=\sevenbf
   \scriptscriptfont\bffam=\fivebf
  \def\tt{\fam\ttfam\tentt}%
  \textfont\ttfam=\tentt
  \tt \ttglue=.5em plus.25em minus.15em
  \normalbaselineskip=12pt
  \let\sc=\eightrm
  \let\big=\tenbig
  \setbox\strutbox=\hbox{\vrule height8.5pt depth3.5pt width\z@}%
  \normalbaselines\rm}

\def\ninepoint{\def\rm{\fam0\ninerm}%
  \textfont0=\ninerm \scriptfont0=\sixrm \scriptscriptfont0=\fiverm
  \textfont1=\ninei \scriptfont1=\sixi \scriptscriptfont1=\fivei
  \textfont2=\ninesy \scriptfont2=\sixsy \scriptscriptfont2=\fivesy
  \textfont3=\tenex \scriptfont3=\tenex \scriptscriptfont3=\tenex
  \def\it{\fam\itfam\nineit}%
  \textfont\itfam=\nineit
  \def\sl{\fam\slfam\ninesl}%
  \textfont\slfam=\ninesl
  \def\bf{\fam\bffam\ninebf}%
  \textfont\bffam=\ninebf \scriptfont\bffam=\sixbf
   \scriptscriptfont\bffam=\fivebf
  \def\tt{\fam\ttfam\ninett}%
  \textfont\ttfam=\ninett
  \tt \ttglue=.5em plus.25em minus.15em
  \normalbaselineskip=11pt
  \let\sc=\sevenrm
  \let\big=\ninebig
  \setbox\strutbox=\hbox{\vrule height8pt depth3pt width\z@}%
  \normalbaselines\rm}

\def\eightpoint{\def\rm{\fam0\eightrm}%
  \textfont0=\eightrm \scriptfont0=\sixrm \scriptscriptfont0=\fiverm
  \textfont1=\eighti \scriptfont1=\sixi \scriptscriptfont1=\fivei
  \textfont2=\eightsy \scriptfont2=\sixsy \scriptscriptfont2=\fivesy
  \textfont3=\tenex \scriptfont3=\tenex \scriptscriptfont3=\tenex
  \def\it{\fam\itfam\eightit}%
  \textfont\itfam=\eightit
  \def\sl{\fam\slfam\eightsl}%
  \textfont\slfam=\eightsl
  \def\bf{\fam\bffam\eightbf}%
  \textfont\bffam=\eightbf \scriptfont\bffam=\sixbf
   \scriptscriptfont\bffam=\fivebf
  \def\tt{\fam\ttfam\eighttt}%
  \textfont\ttfam=\eighttt
  \tt \ttglue=.5em plus.25em minus.15em
  \normalbaselineskip=9pt
  \let\sc=\sixrm
  \let\big=\eightbig
  \setbox\strutbox=\hbox{\vrule height7pt depth2pt width\z@}%
  \normalbaselines\rm}

\input tikz
\usetikzlibrary{arrows.meta,calc,cd,decorations.pathmorphing}

\def\ref #1{{\bf [#1]}} 

\def\br{\hbox{\bf R}}
\def\bc{\hbox{\bf C}}
\def\bz{\hbox{\bf Z}}

\newcount\listcount
\def\smallist{\advance\listcount by 1\item{\bf \the\listcount. }} 
\newcount\figurecount
\global\figurecount=1
\def\fig{{\bf Fig. \the\figurecount:\ } {\global\advance\figurecount by 1}}

\magnification=1200

\hoffset-1.4truecm
\voffset=-0.5truecm
\hsize =18.25truecm 
\vsize=23.34truecm

\font\entryfont=cmbx10 scaled\magstep1
\font\contentsfont=cmbx10

\font\eufscript=eufm10

\newbox\smallstrutbox
\setbox\smallstrutbox=\hbox{\vrule height2pt depth0.5pt width\z@}
\def\smallstrut{\relax\ifmmode\copy\smallstrutbox\else\unhcopy\smallstrutbox\fi}

\newbox\bigstrutbox
\setbox\bigstrutbox=\hbox{\vrule height10pt depth3.5pt width\z@}
\def\bigstrut{\relax\ifmmode\copy\bigstrutbox\else\unhcopy\bigstrutbox\fi}

\newbox\evenbiggerstrutbox
\setbox\evenbiggerstrutbox=\hbox{\vrule height15pt depth3.5pt width\z@}
\def\evenbiggerstrut{\relax\ifmmode\copy\bigstrutbox\else\unhcopy\evenbiggerstrutbox\fi}

\def\directlimit#1{\raise 7truept\hbox{$\matrix{ \lower 5truept\hbox{$\scriptscriptstyle \rightarrow\;$}\cr  #1 \cr}$}}
\def\inverselimit#1{\raise 7truept\hbox{$\matrix{ \lower 5truept\hbox{$\scriptscriptstyle \leftarrow\;$}\cr  #1 \cr}$}}
\def\circaccent#1{\raise4truept\hbox{${\scriptscriptstyle \circ\atop \textstyle #1}$}}


\def\propersubset{{\scriptstyle \mathrel{\lower6truept\hbox{$\subset$}\atop \not=}}}
\def\propersupset{{\scriptstyle \mathrel{\lower6truept\hbox{$\supset$}\atop\not=}}}

\def\mapright#1{\smash{\mathop{\longrightarrow}\limits^{#1}}}

\def\Nearrow#1{\raise 4truept \hbox{$\scriptstyle#1$}\hskip-7truept\nearrow}
\def\Nearrowdown#1{\nearrow\hskip-7truept\lower 4truept \hbox{$\scriptstyle#1$}}
\def\Nwarrow#1{\raise 4truept \hbox{$\scriptstyle#1$}\,\hskip-5truept\nwarrow}
\def\Searrow#1{\lower 4truept \hbox{$\scriptstyle#1$}\hskip-7truept\searrow}
\def\Searrowup#1{\searrow\hskip-7truept\raise 4truept \hbox{$\scriptstyle#1$}}

\def\innprod#1#2{\left\langle#1,#2\right\rangle}

\def\conh#1{conh\,(#1)}


\def\cas2#1{{\rm cas}_2{(#1)}}


\def\det{det\,}

\def\gothic#1{\hbox{\eufscript #1}}

\def\longbar#1{\setbox1=\hbox{$#1$}
\setbox2=\vbox{\hrule width 0.8\wd1}
\raise0.5\ht1\hbox{${\lower\dp1\box2}\atop\box1$}}  
\def\mediumbar#1{\setbox1=\hbox{$#1$}
\setbox2=\vbox{\hrule width 0.6\wd1}
\raise0.5\ht1\hbox{${\lower\dp1\box2}\atop\box1$}}  

\def\lo\bar w ar#1{\setbox1=\hbox{$#1\kern-1.5truept$}
\setbox2=\vbox{\hrule width 0.85\wd1 }
\raise0.4\ht1\hbox{${\lower\dp1\box2}\atop\raise2pt\box1$}}

\def\scriptl#1{\hbox{\eufscript #1}}

\def\Rad{{\rm Radon\kern 0.1em}}

\def\2one{{\rm II}_1}

\long\def\omit#1{}

\definecolor{colour1}{rgb}{0.2,0.2,0.6} 
\definecolor{colour2}{rgb}{0.01,0.28,1.0} 
\definecolor{colour3}{rgb}{0.4,0.6,0.8} 
\definecolor{colour4}{rgb}{0.54,0.17,0.89} 
\definecolor{colour5}{rgb}{0.16,0.32,0.75} 
\definecolor{colour6}{rgb}{0,0,0.5} 
\definecolor{colour7}{rgb}{0,0.13,0.28}
\definecolor{wheat}{rgb}{0.96, 0.87, 0.7}
\definecolor{yellow-green}{rgb}{0.6, 0.8, 0.2}
\definecolor{yaleblue}{rgb}{0.06, 0.3, 0.57}
\definecolor{skyblue}{rgb}{0.53, 0.81, 0.92}
\definecolor{spirodiscoball}{rgb}{0.06, 0.75, 0.99}
\definecolor{upforestgreen}{rgb}{0.0, 0.27, 0.13}
\definecolor{sunset}{rgb}{0.98, 0.84, 0.65}
\definecolor{sunglow}{rgb}{1.0, 0.8, 0.2}
\definecolor{persianorange}{rgb}{0.85, 0.56, 0.35}
\definecolor{salmon}{rgb}{1.0, 0.55, 0.41}
\definecolor{goldenyellow}{rgb}{1.0, 0.87, 0.0}
\definecolor{fawn}{rgb}{0.9, 0.67, 0.44}
\definecolor{darktangerine}{rgb}{1.0, 0.66, 0.07}
\definecolor{darkorange}{rgb}{1.0, 0.55, 0.0}
\definecolor{cobalt}{rgb}{0.0, 0.28, 0.67}
\definecolor{ao(english)}{rgb}{0.0, 0.5, 0.0}
\definecolor{applegreen}{rgb}{0.55, 0.71, 0.0}
\definecolor{burntorange}{rgb}{0.8, 0.33, 0.0}
\definecolor{cadmiumorange}{rgb}{0.93, 0.53, 0.18}
\definecolor{orange(colorwheel)}{rgb}{1.0, 0.5, 0.0}
\definecolor{orange-red}{rgb}{1.0, 0.27, 0.0}


\catcode`\#=12
\catcode`\%=6
\newcount\entrycounter

\def\entry
\def\subheading{}
\def\subsubheading{}
\def\heading{
\global\advance\entrycounter by 1
\global\referencenumber=1
\global\appendixnumber=0
\global\sectioncounter=0
\global\subsectioncounter=0
\global\footcount=1
\par\vskip0.5\baselineskip\noindent
\xrdef{entrylabel\the\entrycounter}
{\entryfont 
}
\write\entrylist{{\the\entrycounter}{
\write\hreflist{{\the\entrycounter}{
\write\contents{
\par\noindent
\expandafter\string\csname href\endcsname{#entrylabel\the\entrycounter}
{\contentsfont 
}

\newcount\definitioncounter
\long\def\definition
\global\advance\definitioncounter by 1
\par\vskip 0.55\baselineskip  minus 0.55\baselineskip\noindent
\xrdef{deflabel\the\definitioncounter}
{\bf  Definition}{ (
\par\penalty 10000\vskip 0.4\baselineskip  minus 0.4\baselineskip\noindent
\write\contents{\par
\expandafter\string\csname href\endcsname{#deflabel\the\definitioncounter}{\expandafter\string\csname color\endcsname {orange(colorwheel)}
{\contentsfont  Definition  \rm 
}

\newcount\remarkscounter
\long\def\remarks
\global\advance\remarkscounter by 1
\par\vskip 0.55\baselineskip  minus 0.55\baselineskip\noindent 
\xrdef{remarkslabel\the\remarkscounter}
{\bf Remarks}{ \it 
\par\penalty 10000\vskip 0.4\baselineskip  minus 0.4\baselineskip\noindent
\write\contents{\par
\expandafter\string\csname href\endcsname{#remarkslabel\the\remarkscounter}{\expandafter\string\csname color\endcsname {cobalt}
{\contentsfont Remarks  \rm 
}

\newcount\summarycounter
\long\def\summary
\global\advance\summarycounter by 1
\par\vskip 0.55\baselineskip  minus 0.55\baselineskip\noindent 
\xrdef{summarylabel\the\summarycounter}
{\bf Summary}{ \it 
\par\penalty 10000\vskip 0.4\baselineskip  minus 0.4\baselineskip\noindent
\write\contents{\par
\expandafter\string\csname href\endcsname{#summarylabel\the\summarycounter}{\expandafter\string\csname color\endcsname {cobalt}
{\contentsfont Summary  \rm 
}

\newcount\examplecounter
\long\def\example
\global\advance\examplecounter by 1
\par\vskip 0.55\baselineskip  minus 0.55\baselineskip\noindent 
\xrdef{examplelabel\the\examplecounter}
{\bf Example}{ \it 
\par\penalty 10000\vskip 0.4\baselineskip  minus 0.4\baselineskip\noindent
\write\contents{\par
\expandafter\string\csname href\endcsname{#examplelabel\the\examplecounter}{\expandafter\string\csname color\endcsname {ao(english)}
{\contentsfont Example  \rm 
}

\newcount\theoremcounter
\definecolor{theoremcolor}{rgb}{0.98,0.94,0.75} 
\long\def\theorem
\global\advance\theoremcounter by 1
\par\vskip 0.55\baselineskip  minus 0.55\baselineskip\noindent
\xrdef{theoremlabel\the\theoremcounter}
{\bf Theorem} (
\par\penalty 10000\vskip 0.4\baselineskip  minus 0.4\baselineskip\noindent
\write\contents{\par
\expandafter\string\csname href\endcsname{#theoremlabel\the\theoremcounter}{\expandafter\string\csname color\endcsname {red}
{\contentsfont  Theorem  \rm 
}

\newcount\sectionlabelcounter
\newcount\sectioncounter
\long\def\newsection
\def\subheading{
\def\subsubheading{}
 \global\subsectioncounter=0
\global\advance\sectioncounter by 1
\global\advance\sectionlabelcounter by 1
\edef\try{\the\sectioncounter}
\par \vskip 0.6\baselineskip 
\noindent
\xrdef{seclabel\the\sectionlabelcounter}
{\bf $\S$ \the\sectioncounter.    
\par\penalty10000
\noindent
{\let\the=0\edef\next{
\write\contents{\par
\expandafter\string\csname href\endcsname{#seclabel\the\sectionlabelcounter}
{\contentsfont $\S$ \try.  
\par\vskip0.5\baselineskip}
}
\next}
\par\vskip0.15\baselineskip
}

\newcount\subsectionlabelcounter
\newcount\subsectioncounter
\definecolor{subsectioncolor}{rgb}{0.98,0.94,0.75} 
\long\def\newsubsection
\def\subsubheading{
\global\advance\subsectioncounter by 1
\global\advance\subsectionlabelcounter by 1
\edef\try{\the\sectioncounter.\the\subsectioncounter}
\par\vskip 0.55\baselineskip  minus 0.55\baselineskip
\noindent
\xrdef{subsectionlabel\the\subsectionlabelcounter}
{\bf $\S\S$ \the\sectioncounter.\the\subsectioncounter\  
\par\penalty 10000\vskip 0.4\baselineskip  minus 0.4\baselineskip\noindent
{\let\the=0\edef\next{
\write\contents{\par
\expandafter\string\csname href\endcsname{#subsectionlabel\the\subsectionlabelcounter}
{\contentsfont $\S\S$\    \try\   \rm 
\par\vskip0.5\baselineskip}
}
\next}
}

\newcount\romanlistcount
\def\makenewromanlist{\global\romanlistcount=1}
\def\romanlist{\item{\rm (\romannumeral\romanlistcount)} \global\advance\romanlistcount by 1}

\newcount\minisectionlistcounter
\newcount\minisectionlistlabelcounter
\long\def\minisectionlist
\global\advance\minisectionlistcounter by 1
\global\advance\minisectionlistlabelcounter by 1
\par\vskip 0.55\baselineskip  minus 0.55\baselineskip\noindent 
\xrdef{minisectionlistlabel\the\minisectionlistlabelcounter}
{\item{\rm (\romannumeral\minisectionlistcounter)}} {\bf 
\par\penalty 10000\item{}
{\let\the=0\edef\next{
\write\contents{\par
\expandafter\string\csname href\endcsname{#minisectionlistlabel\the\minisectionlistlabelcounter}{\expandafter\string\csname color\endcsname {red}
{\contentsfont {$\bullet$} \rm (\romannumeral\minisectionlistcounter) \rm 
}
\next}
}

\catcode`\%=14\catcode`\#=6

\newif\ifdorefdebug \dorefdebugtrue 
\def\Dref@log#1{\ifdorefdebug \immediate\write16{#1}\fi}

\def\Dref@normalize{%
  \ifhmode
    \skip0=\lastskip
    \Dref@log{[doref] lastskip=\the\skip0}%
    \dimen0=\fontdimen2\font \multiply\dimen0 by 2
    \ifdim\skip0>\dimen0
      \unskip
      \hskip \fontdimen2\font plus \fontdimen3\font minus \fontdimen4\font
    \fi
  \fi
}

\long\def\dorefentry#1#2{%
  \ifvmode \noindent\fi       
  \Dref@normalize             
  \hlstart{name}{}{entrylabel#1}#2\hlend
}


\long\def\corollary#1#2{
\par\vskip 0.55\baselineskip  minus 0.55\baselineskip\noindent 
{\bf Corollary} (#1){ \it #2}
\par\penalty 10000\vskip 0.4\baselineskip  minus 0.4\baselineskip\noindent}

\long\def\lemma#1#2{
\par\vskip 0.55\baselineskip  minus 0.55\baselineskip\noindent 
{\bf Lemma} (#1){ \it #2}
\par\penalty 10000\vskip 0.4\baselineskip  minus 0.4\baselineskip\noindent}

\long\def\conjecture#1#2{
\par\vskip 0.55\baselineskip  minus 0.55\baselineskip\noindent 
{\bf Conjecture} (#1){ \it #2}
\par\penalty 10000\vskip 0.4\baselineskip  minus 0.4\baselineskip\noindent}

\long\def\proposition#1#2{
\par\vskip 0.55\baselineskip  minus 0.55\baselineskip\noindent 
{\bf Proposition} (#1){ \it #2}
\par\penalty 10000\vskip 0.4\baselineskip  minus 0.4\baselineskip\noindent}

\long\def\axiom#1#2{
\par\vskip 0.55\baselineskip  minus 0.55\baselineskip\noindent 
{\bf Axiom}{ (#1) }{ \it #2}
\par\penalty 10000\vskip 0.4\baselineskip  minus 0.4\baselineskip\noindent}

\def\proof{
\par\vskip 0.55\baselineskip  minus 0.55\baselineskip\noindent 
{\bf Proof}
\par\penalty 10000\vskip 0.4\baselineskip  minus 0.4\baselineskip\noindent}

\def\proofwithtitle#1{
\par\vskip 0.55\baselineskip  minus 0.55\baselineskip\noindent 
{\bf Proof}{ (\it #1) }
\par\penalty 10000\vskip 0.4\baselineskip  minus 0.4\baselineskip\noindent}

\long\def\description#1#2{
\par\vskip 0.55\baselineskip  minus 0.55\baselineskip\noindent 
{\bf Description}{ (#1) }{ \it #2}
\par\penalty 10000\vskip 0.4\baselineskip  minus 0.4\baselineskip\noindent}

\def\arXiv#1{\href{https://arxiv.org/pdf/#1}{#1}}

\def\pdfimage#1, #2{  
\pdfximage height #1 {#2.pdf}
\pdfrefximage \pdflastximage
}

\newbox\colouredfootbox
\newcount\footcount
\footcount=1

\long\def\foot#1{\footnote{$^{\,\the\footcount}$}{
\eightpoint\noindent  #1}    
\global\advance\footcount by 1}

\def\heading{}               
\def\subheading{}          
\def\subsubheading{}    
\output={\global\advance\pageno by 1\shipout \vbox to \vsize{\boxmaxdepth=\maxdepth 
\ifvoid\topins\else\unvbox\topins\fi
{\null\par\noindent\bf \color {blue}\heading\par\noindent\color{cobalt}\ninepoint\bf\subheading\par\noindent\color{red}\eightpoint\bf\subsubheading\par\par\vskip0.25\baselineskip plus 0.25\baselineskip minus 0.125\baselineskip}
\dimen0=\dp255 \unvbox255
\ifvoid\footins\else
\color {black} 
    \vskip\skip\footins\unvbox\footins\fi
    \pagebottom}
\ifnum\outputpenalty>-20000 \else\dosupereject\fi}

\global\pageno=0
\def\maintextpagebottom{\vbox{\color {black} 
b\hbox{\null}\hbox to \hsize{\hfill{\rm \the\pageno}\hfill} \hbox{{\eightpoint \copyright\  C. Nash, \the\year.}}}}
\def\romanpagebottom{\vbox{\hbox{\null}\hbox to \hsize{\hfill{\rm \romannumeral\pageno}\hfill} \hbox{{\eightpoint \copyright\  C. Nash, \the\year.}}}}


\def\fthreemetric{g^{\scriptscriptstyle F_3}}
\def\fthreeqmetric{g^{\scriptscriptstyle F_3/T}}

\font\midsy=cmsy10 at 6.4pt
\def\middot{\mathbin{\raise0.9pt\hbox{\midsy\char15 }}} 

\def\pagebottom{\centerline{\bf \color{black}  \the\pageno}}
\font\bfbigger=cmbx10 scaled\magstep1

\newcount\sectioncounter
\def\startsection#1{
\global\subsectioncounter=0
\advance\sectioncounter by 1
\par \vskip 0.6\baselineskip 
\noindent
{\bf $\S$ \the\sectioncounter. #1}
\par\penalty10000\vskip0.6\baselineskip
\noindent
}

\newcount\subsectioncounter
\def\startsubsection#1{
\advance\subsectioncounter by 1
\par \vskip 0.6\baselineskip 
\noindent
{\bf $\S$ \the\sectioncounter.\the\subsectioncounter\   #1}
\par\vskip0.6\baselineskip
\noindent
}

\newcount\refcounter
\def\refentry{
\advance\refcounter by 1
\par\vskip0.25\baselineskip\noindent {\bf \the\refcounter. }}

\def\startappendix{
\par \vskip 0.6\baselineskip 
\noindent
\centerline{\bfbigger Appendix}}
\par\vskip0.6\baselineskip
\noindent

\def\appendixsection#1{
\par \vskip 0.6\baselineskip 
\noindent
{\bf $\S$ #1}
\par\vskip0.6\baselineskip
\noindent
}

\def\startreferences{
\par \vskip 0.6\baselineskip 
\noindent
\centerline{\bfbigger References}
\par\vskip0.6\baselineskip
\noindent
}

\def\beginabstract{\par\vskip0.8\baselineskip 
                 plus 0.8\baselineskip
                 minus 0.8\baselineskip
\ninepoint\rm\narrower\narrower}
\def\endabstract{\par\vskip0.8\baselineskip 
                 plus 0.8\baselineskip
                 minus 0.8\baselineskip
                }

\hsize=6in
\leftmargin=0.7in
\overfullrule =0pt
\color{black}{
\centerline{\bfbigger The Geometry of the CKM matrix,}
\centerline{\bfbigger the Standard Model and  RG fixed points}
\par\vskip1.5\baselineskip\noindent

\centerline{Brian~P.~Dolan\footnote{*}{bdolan@stp.dias.ie}
and Charles Nash\footnote{$^\dagger$}{cnash@stp.dias.ie} }

\medskip

 \centerline {\it School of Theoretical Physics,\ Dublin\ Institute\ for\ Advanced\ Studies,}
\centerline{ \it 10 Burlington Rd,\  Dublin,\ Ireland}
\centerline{and}
\centerline{\it Department of Physics,\ Maynooth\ University,\ Ireland}

\par\vskip1.5\baselineskip\noindent

{\beginabstract
 We investigate the geometry of the { Fermion} mixing matrix in the  Standard Model.  The principal structure studied is the CKM matrix,  $V$, viewed as an element of the double coset space $M$, where 
 $M=T\backslash SU(3)/T$, 
and  $T=U(1)\times U(1)$.   The space $M$ is given  a Riemannian metric, and  has both   point singularities, which it is shown correspond to the known {  renormalization} group fixed points, and line singularities connecting the fixed points.
A key discrete object inducing many of these   properties is the Weyl group, $W$, of $SU(3)$, $W$ being the symmetric group $S_3$,
which appears in the guise of   the vertices of a hexagon in $V$.

\endabstract}

\rightline{DIAS-STP-26-21}

\startsection{Introduction}
In this paper we describe the basic geometry and  topology underlying the Standard Model which stems from the CKM matrix, the quark mixing matrix of the Standard Model \ref{1,2}. It
contains the main information on the weak interaction properties of the three quark generations, including that of CP violation.  
It  is a $3\times 3$ unitary matrix  $V$
and, in the conventional choice, has unit determinant, so that $V\in SU(3)$. 
\par
One also  commonly  writes
$$V=
\left[
\matrix{
 c_{12}\, c_{13} & c_{13} \,s_{12} & e^{-i \delta } s_{13} \cr
 -c_{23}\, s_{12}-e^{i \delta } c_{12}\,  s_{13} \, s_{23} &
   c_{12}\, c_{23}-e^{i \delta } s_{12}\, s_{13} \,s_{23}
   & c_{13} \, s_{23} \cr
 s_{12}\, s_{23}- e^{i \delta } c_{12}\, c_{23}\, s_{13} &
   -c_{12}\, s_{23}-e^{i \delta }  c_{23} \, s_{12}\, s_{13}
   & c_{13}\, c_{23} \cr
}
\right]
$$
where
$$c_{ij}=\cos(\theta_{ij})\quad s_{ij}=\sin(\theta_{ij}),\quad\hbox{for appropriate values of $(i,j)$,}$$
$\theta_{23}, \theta_{13},\theta_{12}$ are the quark mixing angles, $0\le \theta_{ij}\le {\pi \over 2}$, and 
$\delta$ is the  usual CP violating parameter,  $0 \le \delta < 2 \pi$.
\par
For what follows, one  should now observe that, though $V\in SU(3)$---which would allow it to
depend on up to $8$ real parameters---$V$ only depends on the $3$ mixing angles and the
CP  violating parameter $\delta$.
This comes about because, if one starts with a general 8 parameter $SU(3)$ matrix, then  acting on the right, or the left,  by
the maximal torus $T=U(1)\times U(1)$ of $SU(3)$  only changes $2+2$ unphysical quark phases. 
Hence  the set of CKM matrices is the {$4$ dimensional  space}
$$T\backslash SU(3)/T$$
which we see   is a  {\it double coset space}; and we denote it by $M$.
So each element   of
$$M=T\backslash SU(3)/T$$
contains the physical information of some CKM matrix. 
\par
A $6$ dimensional space is also of importance here it is the {\it single coset space} $SU(3)/T$. This space is called a  complete  flag
manifold, and is often written as  $F_3$,  thus, declaring  that the second  quotient acts  on the { left},  yields  
$$F_3={SU(3)\over T}, \quad
                 M={F_3\over T}.$$
$F_3$ is a complex $3$-manifold  and is also K\"ahler; still more, it admits a K\"ahler-Einstein metric as we shall see below.          
\par
Fixed points of the 1-loop RG running of $V$ in the Standard Model were found in
\ref{3}: there are 6 of them and they { form} a finite group, the Weyl group of $SU(3)$.  In \ref{4} a proof was given that these remain fixed points of the RG flow to all orders in perturbation theory; furthermore a proof was given in \ref{4} that, if it is assumed that the RG flow on $M$ lifts to a flow on $F_3$ that commutes with the left action of $T$, then these are fixed points of the complete RG flow, to all orders in perturbation theory and even non-perturbatively.

Now, though $F_3$ is a manifold the quotient $F_3/T$ is not a manifold, because it is not differentiable: $T$ acts on  $F_3$ with fixed points. In fact these fixed points of this action coincide with the RG fixed points \ref{5}.
The fixed points of action of $T$ on $F_3$  are easy to identify:  one  proceeds as follows. Let the   coset
$$[gT]\in F_3 $$
be a fixed point---i.e. for some $t\in T$
$$t\middot  [g T]=[g T].$$
By definition
$$t\middot  [g T]=[t g T].$$
Hence
$$t g T= g t^\prime,\;t^\prime\in T.$$
Thus for some $t^{\prime\prime}\in T$  one has
$$\eqalign{t g t^{\prime\prime}&=g t^\prime\cr
\Rightarrow g^{-1} t g t^{\prime\prime}&=t^\prime\cr
\Rightarrow g^{-1} t g&\in T\cr
\Rightarrow g&\in N(T).}
$$
Thus the fixed point set of $F_3$ under this $T$ action is the set $F$, where
$$F=\{ [g T]\mid g\in N(T) \}={N(T)\over T}=W$$
where $W$ is the Weyl group of $SU(3)$.
\par
We already know that $W=S_3$, where $S_3$ denotes    the appropriate symmetric
group---so there are exactly six fixed points and these  form a hexagon.  
We can check this number  against  the Atiyah-Bott fixed point theorem, which says
that the number $\vert F \vert$ of isolated fixed points,  is given by
$$\vert F \vert =\chi(F_3)$$
and, since $\chi(F_3)=6$, this check  succeeds.  
\par
Hence we now know that $M$ is not a manifold because of its  six singular points---some authors use the term {\it manifold
with corners}.  Actually the topology  of $M$ has been uncovered by Buchstaber and Terzi\'c in reference \ref{6}, and it turns  out that
$M$ is homeomorphic to $S^4$.
\par
A short summary of  flows on $F_3$ and $F_3/T$  is as follows.   
Let $g(t)$, $t\in\br$  denote a one parameter curve on
$$SU(3)$$
this induces a curve on $F_3$ with coset  
$$[g(t)T]\in F_3$$
and a further curve on $F_3/T$ with corresponding  coset  
$$[Tg(t)T]\in F_3/T$$
and so one has the quotient map
$$\eqalign{&Q:F_3\longrightarrow F_3/T\cr
                     &\quad\, [g(t)T]\longmapsto [Tg(t)T].}
                     $$
One must not forget   that  $F_3$ is  manifold so the hexagon fixed points are regular in $F_3$
whereas their orbits, as points of $F_3/T$, are singular, and isolated.
\par
Geometrically speaking a flow that begins at a vertex of the hexagon must end at another vertex;  one can
contemplate various ways that this might happen; we shall return to this later. 
\par
Physically  speaking this would depend on the correct alternation of repulsive and attractive properties of each vertex.
These  properties are
determined by the   appropriate Jacobian  and, before going further,  we must introduce  some relevant notation:
we write 
$$F_3/T=M=S^4$$
and, for $t\in \br$,  let
$$\lambda(t)=(\lambda^1,\dots,\lambda^4)$$
be an RG  flow on $M$, so that  
$$\lambda(t)  = [Tg(t)T]. $$
This flow has a critical point at $t=t_p$ when   
$$\left.{d\lambda(t)\over dt}\right\vert_{t=t_p}=0$$ 
 $\lambda(t_p)$ being the  critical point on $M$.   
More explicitly  the integral $ \lambda(t)$ has a generating vector field $\Lambda\in T_\lambda$ given by 
$$\Lambda=\Lambda^1{\partial\over \partial \lambda^1}+\cdots +\Lambda^4{\partial\over\partial \lambda^4}$$
yielding  
$${d \lambda(t)\over dt}=\Lambda(\lambda(t)).$$
Thus $\Lambda$ is just a $4$ component $\beta$ function; so one can write 
$$\beta^i(t)=\Lambda^i(\lambda(t)),\quad i=1,\ldots,4.
                                         $$
                    The Jacobian needed to determine the stability properties of the critical points is 
$$J=\left[{\partial \beta^i(\lambda(t)) \over \partial \lambda^j} \right]_{4\times 4}$$
and for this  one needs  the eigenvalues of $J$.
\par
This is studied  at 1-loop in \ref{4}, including  a search for  the fixed points in perturbation  theory, and, as expected from the geometry,
these fixed points are present, and unchanged, to  all orders  in perturbation  theory.
\par
In the next two sections  we shall explain how $M$ originates, introduce and construct a natural Riemannian metric on $M$,
and relate all  this to the $ SU(3)$ Weyl group $W$
and the CKM matrix $V$;   a  subsequent section will treat singular behaviour of the metric on $M$. 
\startsection{Moment maps and  the hexagon}
Moment maps usually arise when a symplectic  manifold ${\cal M}$  is  acted on  by a Lie group  $G$, whose  action  preserves the
symplectic form of ${\cal M}$.  If ${\gothic g}$ is the Lie algebra of $G$, and we denote the  moment map by $\mu$, it
takes the form
$$\mu:{\cal M}\longrightarrow {\gothic g^*}$$
where ${\gothic g^*}$ is the dual of ${\gothic g}$. 
\par
The moment map $\mu$ can be regarded as encoding a set of Hamiltonian vector fields on ${\cal M}$:  let ${\cal M}$ have  symplectic form 
$\omega$, and suppose 
$\alpha\in {\gothic g}$ generates a symplectic automorphism so that 
$$L_\alpha \omega=0,\quad L \hbox{ denoting Lie derivative.}$$
Then for  $x\in {\cal M}$,   the function $H_\alpha$ defined by
$$H_\alpha=\innprod{\mu(x)}{\alpha}$$
is a Hamiltonian function satisfying  $d H_\alpha=0$,  so that $H_\alpha$  is conserved. Consequently  
 each $\alpha\in\scriptl{g}$ generates a Hamiltonian flow,  and each $H_\alpha$ is constant on its own integral curve. 
\par
In addition to  the action on ${\cal M}$,  there is always an action on ${\gothic g}^*$---the coadjoint action---and
$\mu$ is required  to
be equivariant, or  compatible, with respect to both actions; also for $G$ compact we can use either  ${\gothic g}$ or ${\gothic g}^*$, since the
Killing form allows the identification  ${\gothic g}\simeq {\gothic g}^*$. In addition there can be a  cohomological  obstruction to equivariance,
but this can  be adjusted to be zero  for $G$ compact.
\par

An adjoint orbit belongs to ${\gothic g}$ ($\simeq {\gothic g}^*$) so,
{  when ${\cal M}$  is a copy of $G/T$,
$\mu$ just becomes an {\it inclusion} ${\cal M}\mapright{i}{\gothic g}$.    }
\par
Next we have the key convexity result---cf. Atiyah \ref{7,8}---that  the orthogonal projection
$\pi$  of an adjoint orbit from ${\gothic g}$ onto
 ${\gothic t}$  is precisely the convex hull of the corresponding  $W$ orbit in ${\gothic t}$. 
Hence one can  form the  composite  map
$$\mu:{\cal M}\mapright{i}{\gothic g}\mapright{\pi} {\gothic t}$$
and,   Atiyah \ref{8} shows that the  image $\mu({\cal M})$ is a convex polytope $P$, and the vertices of $P$ are the images of the fixed points
on ${\cal M}$.  Moreover $P$  is precisely the convex hull of the
Weyl orbit in  ${\gothic t}$,  in  other words 
$$\matrix{\mu({\cal M})=\conh{ Wt}\cr
                                      =P \quad\cr}  $$
where $\conh{X}$ denotes the convex hull of a set $X$. 
\par
Slightly  more explicitly,  for $t\in {\gothic t}$, one has  two orbits: the   adjoint orbit  of $t$ in ${\gothic g}$---this is our
symplectic manifold ${\cal M}$---and the  $W$ orbit of $t$ in  ${\gothic t}$, which we  shall denote by $Wt$. Hence the setup is  
$${\cal M} \subset {\gothic g},\qquad W t\subset {\gothic t}$$
viewing  ${\gothic t}$  as a Euclidean space,   
and the projection $\pi$ now yields   
$$\pi: {\cal M}  \longrightarrow {\gothic t}$$
where
$$\pi({\cal M})=P.$$
A small notational  change can be helpful here: we deploy the  moment map notation $\mu_\pi$, where
$$\mu_\pi=\pi$$
and then our moment map data above appears as
$$\mu_\pi:{\cal M}\longrightarrow {\gothic t},\qquad \mu_\pi({\cal M})=P$$
thus $\mu_\pi$ is  a moment map with Abelian group $T$.  
\par
If $T$ is replaced by a non Abelian group $G$, $\mu({\cal M})$ is not  a convex polytope, but
an important theorem of  Kirwan \ref{9} has shown that a
convex polytope also arises here as follows.
Let  ${\cal M}$ be  symplectic,  $G$ be compact connected and non Abelian, and $\mu$
be  a moment map
 $$\mu:{\cal M}\longrightarrow {\gothic g}^*$$
 compatible with the usual actions on ${\cal M}$ and ${\gothic g}^*$, then there is 
convex polytope $\Delta({\cal M})$ given by a certain intersection, of $\mu({\cal M)})$ with a positive  Weyl Chamber ${\gothic t}_+^*$, namely 
$$\Delta({\cal M})=\mu({\cal M})\cap {\gothic t}_+^*.$$
To see why this result collapses to our Abelian  result $\mu_\pi({\cal M})=P$ when $G$ is the torus $T$, we observe the following:  the Weyl group of a torus is trivial since it has no positive roots so the positive Weyl chamber ${\gothic t}_+^*$ is all of ${\gothic t}$.   Further
$$\mu({\cal M})\subset {\gothic t}^*$$
so
$$\Delta({\cal M})=\mu({\cal M})\cap {\gothic t}_+^*=\mu({\cal M})$$
which is our original polytope $P$.
\par
Now, turning  to our symplectic  manifold of interest, we have  the flag manifold $F_3$ which  is the adjoint orbit, and the hexagon, together with
its interior, which is the polytope $P$.
The  vertices of $P$  are created by the Weyl group  $S_3$ acting on a single point  in $(x,y,z)\in {\bf R}^3$; such a polytope, when 
generated by the $n!$ permutations of the $n$  coordinates
of a single point in ${\bf R}^n$,  is called a {\it permutahedron}; it  clearly lies in a hyperplane in ${\bf R}^n$ and so has  dimension $n-1$;   it
is often denoted by $P {\bf e^{n-1}}$. .
\par
Now move on  from $F_3$ to $F_3/T$ itself,  for which we have recourse to \ref{6}. 
Reference   \ref{6} studies a large class of even dimensional  manifolds which are denoted by $M^{2n}$ and called $(2n,k)$ manifolds,  all
these  $M^{2n}$ possess an effective action of a $k$ torus $T^k$.   They are equipped with moment maps
$$\mu:M^{2n}\longrightarrow {\bf R}^k,\quad \mu M^{2n}=P^k$$
where $P^k$ is  some---not necessarily simple\foot{The vertices of a  simple polytope all have the same valence.}---convex polytope.
\par
There are many $(2n,k)$ manifolds---including the complete flag manifolds---about which we know from \ref{6,7} that
$$M^{2n}=F_{k+1}\Rightarrow \mu F_{k+1}=P {\bf e^k}.$$
Our case just concerns    $F_3$, which  we note in passing is  a $(6,2)$ manifold.  
Moreover our CKM manifold $M$ is $F_3/T$ and,   in \ref{6}, we find the important result that
$$F_3/T\simeq A*B$$
for appropriate spaces $A$ and $B$.  
Here  $A*B$ is the standard notation  for the topological join of two topological  spaces,  cf.  \ref{10};  and it is important to know that
$$\dim A*B=\dim A +\dim B+1.$$
In our case $A$ is the boundary $\partial (P{\bf e^2})$ of $P{\bf e^2}$ which is  a hexagon,   and $B$ is ${\bf CP^1}$.
Hence we have  
$$F_3/T\simeq \partial (P{\bf e^2})*{\bf CP^1}.$$
One also knows that
$$\partial (P{\bf e^2})*{\bf CP^1}\simeq S^4$$
but $\partial (P{\bf e^2})*{\bf CP^1}$ is homeomorphic to $S^1 * S^2$ and
$S^m * S^n$ is homeomorphic to  $S^{m+n+1}$, \ref{10}, 
so finally we have
$$F_3/T\simeq S^4.$$
Observe that  this $S^4$ will not be a manifold, because it is a join with the hexagon $\partial (P{\bf e^2})$, { it is not a differentiable $S^4$, there are singularities.}

\startsection{Metrics and curvatures for  $F_3$ and $F_3/T$ }
Next we wish to study metrical and curvature   data on $F_3$ and $F_3/T$, beginning   with $F_3$.
$F_3$ is a complex $3$-manifold which is also K\"ahler, thus it possesses a K\"ahler metric; this  metric is  obtainable
from a corresponding K\"ahler potential.
\par
We found it  useful to derive  this potential, but, as the derivation has quite a few non trivial steps,
we slim this section  down by relegating  the derivation to an appendix.
\par
The  K\"ahler potential $K(z_1,z_2,z_3)$ for $F_3$ is given by
$$K(z_1, z_2, z_3) = \ln(X_1) + \ln(X_2)$$
where
$$X_1 = 1 + |z_1|^2 + |z_2|^2,\quad
X_2 = 1 + |z_3|^2 + |z_1 z_3-z_2|^2$$
and $(z_1,z_2,z_3)$ are   local complex   coordinates.
\par
The $F_3$ metric $\fthreemetric$ is now given by\foot{Since we are dealing with complex manifolds,
whose coordinates are complex, we shall follow the usual
practice which is that  Greek indices $\alpha,\beta,\gamma,\ldots$, rather  than Latin ones $a,b,c,\ldots$,  are those which
can  be given bars to denote complex conjugates giving $\bar\alpha,\bar\beta$ etc.}   
{ $$\fthreemetric_{\alpha\bar \beta}=\partial_\alpha\partial_{\bar \beta}K(z_1,z_2,z_3).$$}

Setting
$$u = z_1 z_3 - z_2$$
yields   the line element
$$ds^2 = {|dz_1|^2 + |dz_2|^2 \over X_1} - {\left\vert \bar{z}_1 dz_1 + \bar{z}_2 dz_2 \right|^2 \over X_1^2} 
+ { |dz_3|^2 + |du|^2 \over X_2} - {\left| \bar{z}_3 dz_3 + \bar{u} du \right|^2 \over X_2^2}.$$
For calculating the Ricci curvature in this K\"ahler case it is extremely useful to  know  that
$$R_{\alpha\bar\beta}=-\partial_\alpha\partial_{\bar\beta}\ln(g),\quad g=\det[g_{\alpha\bar\beta}].$$
\par
For our metric $\fthreemetric$, using {\it Mathematica}, we find that
$$\det[\fthreemetric]={2\over 
(1 + \vert z_1\vert^2 + \vert z_2\vert ^2)^2 (1 -u\bar z_2 + (z_3 + u \bar z_1 \bar z_3)^2}
$$
and then, on  calculating $R_{\alpha\bar\beta}$,  we discover that
$$R_{\alpha\bar\beta}=2\,\fthreemetric_{\alpha\bar\beta}$$
so that the metric $\fthreemetric$ on $F_3$  is  K\"ahler-Einstein, with a proportionality factor of $2$.  
\par
We shall return to this $F_3$ geometrical data in  the next section but, for now, we want to compute the corresponding data for 
our CKM manifold.
$$M=F_3/T.$$
\par
We denote  the $4D$ metric $F_3/T$ by  $\fthreeqmetric$; its computation requires us to  
 project the $F_3$ metric down to $F_3/T$  by using the  $T$ action.
\par
However, since working with   $\fthreemetric$ gives rather large expressions, we only give the full expression for $\fthreemetric$ in an  appendix.
Here, we shall only use the form of $\fthreemetric$ near a  fixed point{ , which we denote by  $(z_1^*,z_2^*,z_3^*)$}.
\par
The $z_i^*$ are found by noting that, for $t\in T$, under the $T$ action $z_i^*\mapsto t\middot z_i^*$ with
$$t\middot z_i^*=e^{i\alpha_i} z_i^*$$
where these non zero phases $\alpha_i$ are easily obtained. Thus we see that 
$(z_1,z_2,z_3)=(0,0,0)$ is a fixed point.  
\par
Next, if the  $ds^2_{fix}$ is the line element in the fixed point neighbourhood, then  
$$\lim_{z_1,z_2,z_3\rightarrow 0}ds^2\longrightarrow 
ds^2_{fix}=d r_1^2+r_1^2 d\phi_1^2+2 (d r_2^2+r_2^2 d\phi_2^2)+d r_3^2+r_3^2 d\phi_3^2,
$$
with $z_k = r_k e^{i\phi_k}$, and we shall denote the corresponding metric by $\fthreemetric_{fix}$. We note that, as should be the case, this 
metric is {\it non singular}: $F_3$ is everywhere smooth. We now move on to the quotient 
$$F_3/T$$
which will be {\it singular} at the   $F_3$ fixed points. 
\par
Our next task must be  to find the Killing vectors, or isometries, of $\fthreemetric$  that generate the $T$ action.
 Since $T$ is two dimensional, there should be   two Killing vectors, say $K_1$ and $K_2$.
\par
Perusal of the expression for
$\fthreemetric$ above shows that the quantity  
$$
    K = \ln(1 + |z_1|^2 + |z_2|^2) + \ln(1 + |z_3|^2 + |z_1 z_3 - z_2 |^2)
$$
contains the relevant information;  with
$$
    |z_1 z_3 - z_2|^2 = r_1^2 r_3^2 + r_2^2 - 2r_1 r_2 r_3 \cos(\Theta), \quad \Theta = \phi_2 - \phi_1 - \phi_3.
$$
One sees, at once that
$$(\partial_{\phi_2}+\partial_{\phi_1})\Theta=0\quad\hbox{ and } \quad (\partial_{\phi_2}+\partial_{\phi_3})\Theta=0$$
thereby providing the Killing vectors   
$$
    K_1 = \partial_{\phi_1} + \partial_{\phi_2} \quad \hbox{ and }\quad
    K_2 = \partial_{\phi_3} + \partial_{\phi_2}.
  $$
In sum
$$
K_1\,\fthreemetric=0\quad\hbox{ and }\quad 
K_2\,\fthreemetric=0.
$$

Since any  tensor field $h$, say,  will descend to $F_3/T$  if and only if the generators of the $T$ action---the
Killing vectors---satisfy   
$$\left.\matrix{
   {\cal L}_{K_i} h = 0 \quad {\rm (invariance)}\cr
    \quad\iota_{K_i} h = 0 \quad {(\rm horizontality)}\cr
  }\right\},\quad i=1,2
  $$
with ${\cal L}$  denoting  Lie derivative and $\iota$ interior product, then we must now implement these requirements  for  $\fthreemetric_{fix}$.  
\par
We see that   $\fthreemetric$ is invariant (${\cal L}_{K_i} g = 0$) but not yet horizontal. To construct $\fthreeqmetric$, we must restrict
$\fthreemetric $ to the horizontal subspace by requiring that
$$\iota_{K_i} {\fthreemetric} = 0,\quad i=1,2.$$
\par
Carrying this out for  
$ds^2_{fix}$ we compute that 
$$\matrix{
    \iota_{K_1} \fthreemetric_{fix} = r_1^2 d\phi_1 + 2r_2^2 d\phi_2 = 0 \ \Rightarrow \ d\phi_1 = -{2r_2^2\over r_1^2} d\phi_2 \cr
    \iota_{K_2} \fthreemetric_{fix} = r_3^2 d\phi_3 + 2r_2^2 d\phi_2 = 0 \ \Rightarrow \ d\phi_3 = -{2r_2^2\over r_3^2} d\phi_2 \cr
  }
  $$
and substituting  this information  into   $d\Theta$ gives us $d\Theta$ on the horizontal subspace, and this  
yields 
$$
d\Theta = 2r_2^2 C d\phi_2,\quad \hbox{ where } \quad C=  {1\over r_1^2} + {1\over 2 r_2^2} + {1\over r_3^2}.$$
Finally, we must  evaluate  $\fthreemetric$ on the horizontal subspace to give $\fthreeqmetric$. The resulting line element, on $F_3/T$, near the
fixed point, we denote by $ds^2_{fix}(F_3/T)$, and  we obtain   
$$ds^2_{fix}(F_3/T)=dr_1^2+2 dr_2^2+d r_3^2+{1\over C} d\Theta^2. \eqno(1)$$
\par
We anticipate a conical singularity and, to this end,  we 
choose the ray
$$r_1=r, \quad
                        r_2= {r\over \sqrt{2}}, \quad
                        r_3=r.$$
 our line element becomes
$$
3 dr^2 + {r^2\over 3} d\Theta^2
$$
and scaling  by the innocuous factor of 3 yields
$$dr^2+{r^2\over 3^2}d\Theta^2.$$
This shows a circumference of $2\pi r / 3$, giving a  conical deficit  angle of  $2\pi -2\pi/3=4\pi/3$.
\startsection{Riemann and Ricci  tensors for $F_3/T$ near the fixed point}
In this brief section we use curvature data to verify the presence of a singularity   on $F_3/T$.
Let use the notation $(x,y,z,\theta)$ for coordinates on $F_3/T$ so that
$$(x,y,z,\theta)=(r_1,r_2,r_3,\Theta).$$
The Riemann and Ricci tensor give bulky expressions, but, just as satisfactory for our singularity detection is
the Ricci scalar $R_{F_3/T}$, whose computation  yields the more compact formula 
$$
R_{F_3/T}={6 (x^4 (2 y^2 + z^2) + 2 y^2 z^2 (2 y^2 + z^2) + 
   x^2 (4 y^4 + z^4))\over (2 y^2 z^2 + x^2 (2 y^2 + z^2))^2}.
$$
The singularity is at $(x,y,z)=(0,0,0)$; using  the notation of the previous section
we set
$$
(x,y,z)=(r,{r\over \sqrt{2}},r)
$$
thereby obtaining
$$R_{F_3/T}={4\over r^2}$$
which displays the singularity as well as its strength---just for comparison  we add a reminder that the Schwarzschild singularity has
Ricci scalar $R_{\rm Schw}=0$---but this Ricci flatness 
  allows the  Schwarzschild spacetime to be  a solution to the vacuum field equations.
  \par
In order to make a direct comparison between  the  strengths of the two different singularities one should compute,  not the Ricci scalar $R$,
but the {\it Kretschmann scalar} $K$, given  by  $K=R_{\mu\nu\rho\sigma}R^{\mu\nu\rho\sigma}$; then one finds that
    $$
        K_{F_3/T} = {8\over r^4} \quad\hbox{ and }\quad 
        K_{\rm Schw} = {48M^2\over r^6}
 $$
revealing the Schwarzschild  singularity to be  the more singular of the two.
\startsection{The metric $\fthreeqmetric$ in the CKM picture}
The CKM matrix, 
provides us with coordinates for the quotient $F_3/T$, and these are natural physical coordinates coming from  the {\it Standard Model}, so it is of interest
to express $\fthreeqmetric$ in terms of these coordinates. 
\par
{ To achieve this we first} lift $V$ from $F_3/T$ to $F_3$ and create the K\"ahler potential on $F_3$ in our new coordinates.  Then, working in these new  coordinates, we can calculate the  metric on $F_3$ and project it to the quotient. 
\par
Consider  the complexification $SL(3,\bc)$ of $SU(3)$ and its Borel subgroup
$$B\subset SL(3,\bc).$$
$B$  consists of all invertible upper triangular matrices of the form  
$$\left[\matrix{t_1&a&b\cr
           0&t_2 &c\cr
           0&0&t_3\cr}\right],\; t_1 t_2 t_3=1,
$$
with $a,b,c$ and $t_i$ complex.
Next we will need to make use of  the diffeomorphism  
$$F_3={SU(3)\over T}\simeq {SL(3,\bc)\over B}$$
which shows that $F_3$ is  a complex manifold. 
\par
The lifting of $V$ consists of left multiplying $V$ by a torus  element. So, with $t\in T$ given by 
$$t=\left[\matrix{e^{i\alpha}&0&0\cr
                           0&e^{i\beta}&0\cr
                           0&0& e^{-i(\alpha+\beta)}\cr
}\right]$$
we form the coset  
$$[tV]\in F_3.$$
Next we apply $LU$ factorisation\foot{This is so because  if one does an $LU$ factorisation
on any matrix $F$, say, one can write
$$PF=LU$$
where $L$ is unipotent lower triangular, $U$ is upper triangular and $P$ is a possible occurrence of a permutation  matrix.} to  $tV$; and so we write
$$tV=PLU$$
we find---in this particular case, but not all---that $P$ is just the identity;  but $U\in B$ so, coset wise, we can write   
$$[LUB]=[LB]$$ 
and we only  need $L$, which is given by  
$$L  = \left[\matrix{1&0&0\cr
\smallstrut\cr
-{e^{-i(\alpha-\beta)}\over  c_{12}c_{13}}(c_{23} s_{12}+e^{i\delta}c_{12}s_{13} s_{23})&1&0\cr
 \smallstrut\cr
 {e^{-i(2\alpha+\beta)}\over c_{12}c_{13}} (-e^{i\delta}c_{12}c_{23}s_{13}+s_{12}s_{13})& e^{-i(\alpha+2\beta)}{s_{23}\over c_{23}} &1\cr
}\right].
$$
\par
Now we use these new CKM coordinates  to  construct the K\"ahler potential $K$ on $F_3$: to do this
one  defines $L$} by writing
$$L=\left[\matrix{1 & 0 & 0\cr
                      z_1& 1 & 0\cr
                      z_2& z_3 &1\cr             
}\right].$$
We now have the complex coordinates of $F_3$ expressed in terms of
physical quantities: the CKM matrix parameters. Recalling that $K$ is given by
$K=K_1+K_2$ with 
$$K_1=\ln (1+ \vert z_1 \vert^2 +\vert z_2\vert^2)\quad\hbox{ and }\quad  K_2=\ln (1+\vert z_3\vert^2+ \vert z_1 z_3-z_2 \vert^2)$$
we find that $ K_1=-\ln c_{12}^2 c_{13}^2 $ and $ K_2=-\ln c_{13}^2 c_{23}^2$.
\par\noindent
Hence  
$$K=-\ln c_{12}^2 c_{13}^4  c_{23}^2.$$
This CKM variable form of expression for { the} potential $K$ is remarkably compact and simple;  but when one constructs the CKM form of the  metric it
becomes less so.  
\par
The CP violating variable $\delta$ may seem to have  disappeared---indeed it is absent from the K\"ahler potential---but it now resides
in  the $dz_\alpha$. For example, one has, 
$$
\matrix
{
dz_1 = {\partial z_1 \over \partial \theta_{12}} d\theta_{12} 
+ {\partial z_1 \over \partial \theta_{13}} d\theta_{13} 
+ {\partial z_1 \over \partial \theta_{23}} d\theta_{23} 
+ {\partial z_1 \over \partial \delta} d\delta 
+ {\partial z_1 \over \partial \alpha} d\alpha 
+ {\partial z_1 \over \partial \beta} d\beta\cr
\bigstrut{\partial z_1 \over \partial \delta} = -i e^{i(\beta-\alpha)}  \sin\theta_{23} \tan\theta_{13} \,e^{i\delta}.\hfill\cr
}
$$ 
\par
In the CKM coordinates  of the matrix $L$, the fixed point (for which we formerly had, $z_i=r_i e^{i\phi_i}$ and  $z_i \rightarrow 0$)  becomes $\theta_{ij}\rightarrow 0$,   
and one easily calculates that
$$\lim_{\theta_{ij}\rightarrow 0}\left\{\eqalign{
                                                       &z_1\longrightarrow -e^{-i(\alpha-\beta)} \theta_{12}\cr
                                                       &z_2\longrightarrow -e^{-i(2\alpha+\beta+\delta)}\theta_{13}\cr
                                                       &z_3\longrightarrow e^{-i(\alpha+2\beta)}\theta_{23}\cr
                                                       }\right\}\Rightarrow\left\{\eqalign{
r_1\longmapsto \theta_{12},\quad& \quad \phi_1\longmapsto -(\alpha-\beta)\cr
r_2\longmapsto \theta_{13},\quad& \quad \phi_2\longmapsto -(2\alpha+\beta+\delta)\cr
r_3\longmapsto \theta_{23},\quad& \quad \phi_3\longmapsto -(\alpha+2\beta)\cr
}\right.
$$
so that 
$$\eqalign{\lim_{{\theta_{ij}}\rightarrow 0} ds^2 & \rightarrow \ ds^2_{fix}=\cr
& d \theta_{12}^2+\theta_{12}^2  \, (d\alpha-d\beta)^2+2  d\theta_{13}^2
+ 2 \theta_{13}^2\,(2d\alpha+d\beta+d\delta)^2+
                              d\theta_{23}^2+\theta_{23}^2\,(d\alpha+2d\beta)^2.}   
$$
The two previous Killing vectors  now become 
$$
    K_1 = -{2\over 3}\partial_{\alpha} + {1\over 3}\partial_{\beta},\quad
    K_2 = -{1\over 3}\partial_{\alpha} -{1\over 3}\partial_{\beta}
 $$
whose  horizontality  conditions  
$$\iota_{K_1} \fthreemetric(\middot)= 0,\quad
\iota_{K_2}\fthreemetric(\middot)=0$$
produce the two equations
$$\eqalign{\matrix{\theta_{12}^2 (+1 )}\cdot (d\alpha-d\beta)+2  \theta_{13}^2(-1)\cdot (2d\alpha+d\beta+d\delta)+
\theta_{23}^2(-0)\cdot(d\alpha+2d\beta)&=0\cr
 \bigstrut\theta_{12}^2(0)\cdot (d\alpha-d\beta)+2 \theta_{13}^2(-1)\cdot (2d\alpha+d\beta+d\delta)+
\theta_{23}^2(-1)(d\alpha+2d\beta)&=0.\hfill \cr}
$$

In terms of $(x,y,z)=(r_1,r_2,r_3)$ this limit is 
$$(\theta_{12},\theta_{13},\theta_{23})=(x,y,z)$$
and we obtain 
{ $$
d\alpha= -{2\over  3 }{(2  y^2 z^2 +x^2 y^2 )\over 
   (2 y^2 z^2 + 2 x^2 y^2  + x^2 z^2 )}\,d\delta,\quad 
d\beta = {2\over 3}{(  y^2 z^2 -  x^2 y^2 )\over 
  (2 y^2 z^2 + 2 x^2 y^2  +x^2 z^2 )}\,d\delta 
$$}
as well as  the quotient line element  $ds^2_{\scriptscriptstyle (F_3/T)}{\scriptstyle (fix)}$ near the fixed point, which is given by 
$$
ds^2_{\scriptscriptstyle (F_3/T)}{\scriptstyle (fix)}=d \theta_{12}^2 +2  d\theta_{13}^2+ d\theta_{23}^2+
{2  \theta_{12} ^2 \theta_{13}^2 \theta_{23}^2\over 2 \theta_{13}^2 \theta_{23}^2 + 2 \theta_{12}^2 \theta_{13}^2 + \theta_{12}^2\theta_{23}^2} d\delta^2  
$$
and this is identical to equation (1), which used the variables $r_1,r_2,r_3$ and  $\Theta$; something which should not be
surprising; however somewhat interesting is the role  of the Jarlskog invariant, to which we now turn.
\sectioncounter=5
\startsubsection{The Jarlskog invariant}
The Jarlskog invariant ${\cal J}$ is given by 
$${\cal J}= c_{12} \,s_{12} \,c_{23} \,s_{23}\, c_{13}^2 s_{13} \,\sin \delta,$$
so that
 $${\cal J}=0\hbox{ at the fixed point}$$
as well as vanishing  when $\sin\delta=0$.
Near the fixed point, one notes that
$$\lim_{\quad\theta_{ij}\rightarrow 0}{\cal J} \longrightarrow \theta_{12}  \theta_{13}\theta_{23}\sin\delta ={\cal J}_0\sin \delta$$
with the appropriate  definition for ${\cal J}_0$.
Hence 
 on comparing with $ds^2_{\scriptscriptstyle (F_3/T)}{\scriptstyle (fix)}$,
one has 
$$
ds^2_{\scriptscriptstyle (F_3/T)}{\scriptstyle (fix)}=d \theta_{12}^2 +2  d\theta_{13}^2+ d\theta_{23}^2+
{2{\cal J}_0^2 d\delta^2\over 2 (\theta_{13}^2 \theta_{23}^2 + 2 \theta_{12}^2 \theta_{13}^2 + \theta_{12}^2\theta_{23}^2)}.
$$
 
\startsection{Vanishing mixing angles and  singularities on $F_3/T$}
We have seen that there is a  point singularity on $F_3/T$ when all three mixing angles vanish---i.e. 
$$\theta_{12}=\theta_{13}=\theta_{23}=0.$$
But, since the coefficient of $d\delta^2$ in the line element  is
$${2  \theta_{12} ^2 \theta_{13}^2 \theta_{23}^2\over 2 \theta_{13}^2 \theta_{23}^2 + 2 \theta_{12}^2 \theta_{13}^2 + \theta_{12}^2\theta_{23}^2} $$
we should also investigate what happens when a maximum  of two  of the $\theta_{ij}$ vanish, yielding  a space  $M_{vanish}$, say,  of appropriate dimension.  
One can then  examine  the  geometry on $M_{vanish}$: for example, calculate the  intrinsic curvature on  $M_{vanish}$.  
 \par
 A simple stratagem which will allow  this  is to compute the Ricci scalar $R$. So, using the variables $(x,y,z)$ again, where   
 $(\theta_{12},\theta_{13},\theta_{23})=(x,y,z)$, we find that, in the neighbourhood of the fixed point, one has
 $$R={6 (x^4 (2 y^2 + z^2) + 2 y^2 z^2 (2 y^2 + z^2) + 
   x^2 (4 y^4 + z^4))\over (2 y^2 z^2 + x^2 (2 y^2 + z^2))^2}.$$
Next we look at $R$ when precisely one, or two,  of the $(x,y,z)$ vanish.

\medskip

\makenewromanlist
\romanlist {\bf One vanishing mixing angle}
\par
\smallskip
Let us have
$$(x,y,z)=(0,y,z),\quad y\not=0,z\not=0$$
\par
then we find that
$$R_{\vert_{x=0}}={3\over y^2} + {6\over z^2}<\infty$$
\par
 and letting $y$ or $z$ play the role of $z$ we find similar expressions. 
Thus precisely one vanishing $\theta_{ij}$ does not give a singularity.
\par
Pulling back the metric $\fthreeqmetric$ to $M_{vanish}$, and setting $x=0$,
gives an intrinsic metric which is perfectly flat:
$$
ds^2_{intrinsic} = dy^2 + dz^2 \Rightarrow R_{intrinsic} = 0
$$
we see that the $M_{vanish}$ are $2$ dimensional,  and form  smooth, flat faces. 

\medskip

\romanlist{\bf Two vanishing mixing angles}
\par
\smallskip

Next  let
$$(x,y,z)=(0,0,z),\quad z\not=0$$
\par
$M_{vanish}$ is now one dimensional and flat so that
$$
ds^2_{intrinsic} = dz^2$$
and one trivially has
$$R_{intrinsic}=0.$$
However the bulk curvature diverges on $M_{vanish}$:  we easily discover that
$$ \lim_{{x\rightarrow 0\atop y\rightarrow 0}}R \longrightarrow \infty.$$
More precisely 
$$\eqalign{(x,y,z)&=(\epsilon,0,z)\cr
\Rightarrow R&={6\over \epsilon^2} + {6\over z^2}\cr
\Rightarrow R& \hbox{ is singular along an interval  $(0,0,z)$, { $z\in[0,\eta]$.}}\cr
}
$$
\par
More generally  one sees  that there are three {\it line singularities} of $R$ emanating from the fixed point.
In fact, taking account of the other fixed points we see that there will be {\bf  nine} of these singular lines--- they will be mentioned in our next section.

\startsection{The RG flow, the Weyl group and the other fixed points}
\vskip -15pt
\startsubsection{The RG flow lines}
The renormalization group  flow stops and starts at the fixed points and these are the vertices of the cube shown in figure 1 below, with $\delta=0$ or $\pi$.  But only 6 of these points are physically distinct: on the $\theta_{13}={\pi\over 2}$ surface opposite vertices are physically the same. 
\par\vskip0.2\baselineskip
\centerline{\pdfimage0.2\vsize, {Figure1}}
\vskip0.2\baselineskip
\centerline{\bf Figure 1. Fixed points of the RG flow}   
\par\vskip0.2\baselineskip
\noindent A more symmetrical picture is obtained by moving points 5 and  6 down to the same plane as 1 to 4,  and representing the six points  as a hexagon, as in figure 2, known as a permutahedron in the mathematical literature. At 1-loop the edges of the cube are all RG flow lines and the direction of the flow depends on the relative magnitudes of the Yukawa couplings \ref{3}: with the observed Standard Models the IR flow is in the direction of the arrows in figure 2, with the IR fixed point the identity matrix with no mixing.
\par\vskip0.2\baselineskip
\centerline{\pdfimage0.2\vsize, {Figure2}}
\par\vskip0.2\baselineskip
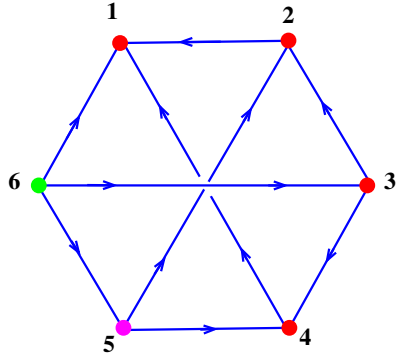
\centerline{\bf Figure 2. The fixed point hexagon and the 1-loop RG flow lines}   
\par\vskip0.25\baselineskip

\noindent
\startsubsection{Fixed point locations and the Weyl  group}
Let us  begin with the fact that
$${G\over T}={G_{\bf c}\over B}.$$
Then, since $B$
consists of upper triangular matrices, including the diagonal,  one can easily verify that { the tangent space}
$$\eqalign{T(G_{\bf c}/B)&={\gothic g}_{\bf c}/{\gothic b}\cr
                                        &=N_\Delta \cr}
                                        $$
where  $N_\Delta$ denotes  the set of strictly  lower triangular  matrices.
So, in an obvious notation,   an  $n_\Delta\in N_\Delta$ looks like
$$n_\Delta=\left[\matrix{0&0&0\cr
                             \middot&0&0\cr
                             \middot&\middot &0\cr} \right]$$ 
which we recognise as  a tangent space point on $F_3$.  If we exponentiate by computing $n_-$ where 
$$n_-=\exp[ n_\Delta]$$
it is straightforward  to check  that $n_-$ is a unipotent lower triangular matrix, so   we can  write 
$$n_-=\left[\matrix{1&0&0\cr
                               z_1&1&0\cr
                               z_2&z_3&1\cr} \right]
                               $$
 and this  is a point on $F_3$. 
Further if we set  $(z_1,z_2,z_3)=(0,0,0)$, then $n_-$ becomes the identity---which is clearly fixed under the torus action---so
this is the   familiar fixed point discussed above.
\par
We shall  refer to $n_-$ as a coordinate patch. However there are five other fixed points
 and they are not  in the coordinate patch $n_-$. We will now use the Weyl group and the Bruhat decomposition to
 find these remaining fixed points,  
\par 
The flag manifold ${G_{\bf c}/B}$ has a CW decomposition indexed by the elements of the Weyl group, and each cell of this decomposition contains
a different fixed point.  This decomposition is called  the 
{\it Bruhat decomposition};  here we supply  a brief description.
\par
The Borel subgroup $B$, and the Weyl group $W$, come together to
give a   remarkable {\it partition}  of $G_{\bf c}$, into {\it double cosets} and this leads at once to a {\it cell decomposition}
of ${G_{\bf c}/ B}$.
The  decomposition takes the form 
$$\eqalign{{G_{\bf c}\over B}&=\bigcup_{w\in W} G^B_w\hfill \cr
                 G^B_w&={G_w\over  B}\quad (G_w=BwB)\hfill \cr
                 G^B_w&\simeq \bc^{l(w)}\hfill \cr}
$$
where $l(w)$ is a certain integer known as the {\it length} of $w$---cf. the appendix.  The diffeomorphism
$$G^B_w\simeq \bc^{l(w)}$$
identifies each $G^B_w$ as a complex cell of  dimension
$l(w)$, and thus of {\it real dimension} $2l(w))$; thereby 
revealing, incidentally,   that the odd Betti numbers of the flag manifold $G/T$ all vanish.
\par
For a cell $G^B_w$ with   $x\in G^B_w$ then, for some matrix $m\in G_{\bf c}$, we have   
$$ x=[w m B].$$
But using $LU $ factorisation, we write $w m=L_wU$ where $L_w$ is unipotent lower   triangular. Hence 
$$\eqalign{
 [wm B]&=[L_wU B]\cr
           &=[L_w B],\quad\hbox{ since }U\subset B\cr},
                             $$
so $ L_w$ gives us  the chart containing the cell  and  fixed point, that  correspond
to  the element $w\in W$.    
\par
We are interested in $F_3$ for which the Weyl group is the symmetric group $S_3$ and  
 $$F_3 \cong SL(3,{\bc })/B$$
 and the top dimensional cell, or big cell,  has dimension $3$. This being the cell $G^B_w$ with 
 $$w=(321)$$
 for which   $l(w)=3$.
\par
When $w=(123)$ (the identity), we  shall conform with our earlier  notation by making   the obvious choice
$m=n_-$ where
$$n_-=\left[\matrix{ 1 & 0 & 0 \cr z_1 & 1 & 0 \cr z_2 & z_3 & 1 \cr }\right] $$
so 
$$L_{(123)}=\left[\matrix{1 & 0 & 0 \cr z_1 & 1 & 0 \cr z_2 & z_3 & 1 \cr }\right]$$
and more generally, $L_w$ is the first  factor in the LU decomposition of $w n_-$, i.e.  
$$ w n_-=L_w U.$$
Now we can display a table of  the chart data for each  $w\in S_3$.
\par
Chart 1 has the fixed point $(0,0,0)$ and the fixed points for the other charts are translates of
$(0,0,0)$ by some Weyl group element $w$. Hence they  occur at $w$ itself, since these translates are the points  
$$w n_-\quad \hbox{ when }z_1=z_2=z_3=0.$$
\medskip
\par\noindent
{\eightpoint \medskip
\par\noindent
{\bf Chart 1} $w = (123)$, $l(w)=0$, cell coords $(0,0,0)$ \hfill 
{\bf Chart 2} $w = (213)$, $l(w)=1$, cell coords $(z_1,0,0)$\qquad\qquad\qquad
$$ w = \left[\matrix{1 & 0 & 0\cr 0 & 1 & 0\cr 0 & 0 & 1\cr}\right],\quad 
   L_w = \left[\matrix{ 1 & 0 & 0 \cr z_1 & 1 & 0 \cr z_2 & z_3 & 1 \cr }\right] \qquad w = \left[\matrix{ 0 & 1 & 0 \cr 1 & 0 & 0 \cr 0 & 0 & 1 \cr }\right], \quad
   L_w = \left[\matrix{ 1 & 0 & 0 \cr {1/ z_1} & 1 & 0 \cr {z_2/ z_1} & z_2 - z_1 z_3 & 1 \cr }\right] $$
\medskip
\par\noindent
{\bf Chart  3} $w = (132)$, $l(w)=1$, cell coords $(0,0,z_3)$ \hfill  {\bf Chart  4} $w = (231)$, $l(w)=2$, cell coords $(z_1,z_2,0)$\qquad\qquad\qquad
$$ w = \left[\matrix{ 1 & 0 & 0 \cr 0 & 0 & 1 \cr 0 & 1 & 0 \cr }\right], \quad
   L_w = \left[\matrix{ 1 & 0 & 0 \cr z_2 & 1 & 0 \cr z_1 & {1/z_3} & 1 \cr }\right] \qquad w = \left[\matrix{ 0 & 1 & 0 \cr 0 & 0 & 1 \cr 1 & 0 & 0 \cr }\right], \quad
   L_w = \left[\matrix{ 1 & 0 & 0 \cr {z_2/z_1} & 1 & 0 \cr {1/z_1} & {1/( z_2-z_1 z_3) } & 1 \cr }\right] $$
\medskip
\par\noindent{\bf Chart 5} 
$w = (312)$, $l(w)=2$, cell coords $(0,z_2,z_3)$ \hfill 
{\bf Chart 6}
$w = (321)$, $l(w)=3$, cell coords  $(z_1,z_2,z_3)$ 
$$ w = \left[\matrix{ 0 & 0 & 1 \cr 1 & 0 & 0 \cr 0 & 1 & 0 \cr }\right], \quad
   L_w = \left[\matrix{ 1 & 0 & 0 \cr {1/z_2} & 1 & 0 \cr {z_1/ z_2} & {(z_1 z_3 - z_2)/ z_3} & 1 \cr }\right] \qquad w = \left[\matrix{ 0 & 0 & 1 \cr 0 & 1 & 0 \cr 1 & 0 & 0 \cr }\right], \quad
   L_w = \left[\matrix{ 1 & 0 & 0 \cr {z_1/z_2} & 1 & 0 \cr {1/  z_2} & {z_3/(z_1 z_3 - z_2)} & 1 \cr }\right] $$
\par\vskip0.5\baselineskip
}
\noindent
{\bf Chart properties}
\par\vskip0.5\baselineskip
Each chart is an open  set of $\bc^3$ with local coordinates $z_1,z_2,z_3$. Each of the six charts has a
single fixed point; if we list the Weyl group elements in the order in which they appear in the chart table,
this fixed point is the Weyl group element
$$w_i$$
for chart $i$.  
A useful
fact,  is that multiplication by $w_j$  translates chart $i$ to chart $k$, where
$$w_j w_i =w_k$$ 
while, if $\Omega^{i,j)}$ denotes the intersection of charts $i$ and $j$, then   $L_{w_i}$ is the simply the transition function
$$\eqalign{&\quad \phi:\Omega^{(i,1)}\longrightarrow \Omega^{(i,1)}\cr
&\qquad\; (z^i_1, z_2^i, z^i_3)\longmapsto (z^1_1,z_2^1,z^1_3)\cr}$$
from chart i coordinates to those of   chart 1 (the identity chart).

\startsection{Conclusions}
The geometry of the CKM matrix inherited from that of the flag manifold $F_3$ has a discrete $S_3$ symmetry which is manifest in the RG flow of the CKM parameters.  The full geometry is rather complicated but it can at least be understood close to the 6 vertices of the associated permutahedron (figure 2), which are fixed points of both the action of $U(1)\times U(1)$ on $F_3$ and of the RG flow.   These fixed points are singular and there are conical singularities there. The Ricci scalar also diverges along the edges of the hexagon, at least close to the vertices, and these edges are flow lines of the RG at 1-loop.

There is thus an intimate connection between the geometry of the flag manifold $SU(3)/U(1) \times U(1)$, the double coset  $U(1)\times U(1) \backslash SU(3) / U(1)\times U(1)$ and the RG flow of the Standard Model.

The considerations here are not only relevant to the quark sector of the Standard Model, but also to the leptonic sector with three sterile right-handed neutrinos,  and the PMNS matrix.  If the right-handed neutrinos are Majorana then are two extra phases in the PMNS matrix, as the left action of $U(1)\times U(1)$ is excluded, the resulting mixing matrix is in $F_3$ itself.  This has a rich geometry, with a complex structure and a K\"ahler metric,
which is much more tractable than that of the double coset, and this geometry will almost certainly  prove useful in studying the RG flow of the leptonic sector, if the neutrinos are Majorana.

\vfill\eject

\startappendix
\appendixsection{A. The explicit $\fthreemetric$ and the   K\"ahler potential construction  for $F_3$}
In $\S\, {\bf 3}$ we assumed the  K\"ahler potential for $F_3$ and gave the form of   its metric near a fixed point,  but we saved clutter by not
giving the full form of the  metric without approximation. Here now   is the full metric
\par

{\hfuzz=\maxdimen 
  \overfullrule=0pt 
  \par\vskip0.5\baselineskip\noindent
$
 \eightpoint \fthreemetric=
\left[\matrix{
 {\vert z_3\vert^2  (\vert z_3\vert^2 +1)\over (-\vert  u \vert^2 +\vert z_3\vert^2 +1)^2}+{\vert z_2\vert^2 +1\over (\vert z_1\vert^2 +\vert z_2\vert^2 +1)^2} & -{z_3 (\vert z_3\vert^2 +1)\over (-\vert  u \vert^2 +\vert z_3\vert^2 +1)^2}-{\bar z_1  z_2\over (\vert z_1\vert^2 +\vert z_2\vert^2 +1)^2} & {z_3 (\bar z_1 +\bar z_2  z_3)\over (\bar  u   z_2+z_3 (\vert z_1\vert^2  \bar z_3 -z_1 \bar z_2 +\bar z_3 )+1)^2} \cr
 -{\bar z_3  (\vert z_3\vert^2 +1)\over (-\vert  u \vert^2 +\vert z_3\vert^2 +1)^2}-{z_1 \bar z_2 \over (\vert z_1\vert^2 +\vert z_2\vert^2 +1)^2} &
 {\vert z_3\vert^2 +1\over (-\vert  u \vert^2 +\vert z_3\vert^2 +1)^2}+{\vert z_1\vert^2 +1\over (\vert z_1\vert^2 +\vert z_2\vert^2 +1)^2} & -{\bar z_1 +\bar z_2  z_3\over (\bar  u   z_2+z_3 (\vert z_1\vert^2  \bar z_3 -z_1 \bar z_2 +\bar z_3 )+1)^2} \cr
 {\bar z_3  (z_1+z_2 \bar z_3 )\over (\bar  u   z_2+z_3 (\vert z_1\vert^2  \bar z_3 -z_1 \bar z_2 +\bar z_3 )+1)^2} & -{z_1+z_2 \bar z_3 \over (\bar  u   z_2+z_3 (\vert z_1\vert^2  \bar z_3 -z_1 \bar z_2 +\bar z_3 )+1)^2} & {\vert z_1\vert^2 +\vert z_2\vert^2 +1\over (\bar  u   z_2+z_3 (\vert z_1\vert^2  \bar z_3 -z_1 \bar z_2 +\bar z_3 )+1)^2} \cr
}\right]
$
\par\vskip1.5\baselineskip
}
Moreover, thus far, we have simply quoted  the K\"ahler potential for our K\"ahler Einstein metric on $F_3$ without constructing it first.
We remedy this shortcoming in this section and, in doing so, it is actually more illuminating to generalise from $F_3$ to $F_n$ and to consider any
K\"ahler metric.
\par
The principal ingredient required for construction of a K\"ahler potential on a K\"ahler manifold $M$
is the second cohomology group $H^2(M;\bz)$; and, for $F_n=G/T$, one knows that $H^2(F_n;\br)$ is generated by the $(n-1)$
classes 
$$[\omega_1],\ldots, [\omega_{n-1}]$$
with $\omega_i$  associated to  a simple root of $G$. Further, for each $i$,   
there is a holomorphic line bundle $L_i$ over $F_n$ whose first Chern  class is given by
$$c_1(L_i)=\left[{\omega_i/2\pi}\right].$$
Finally each $\omega_i$ is a K\"ahler form which satisfies 
$$\omega_i=i\partial \bar \partial K_i$$
so that (by definition)  each $K_i$ is a K\"ahler potential.
\par
Hence we can select a general form $\omega$ from the {\it K\"ahler cone}, where 
$$\eqalign{\omega&=\alpha_1\, \omega_1+\cdots+\alpha_{n-1} \,\omega_{n-1}\cr
                     &\alpha_i>0,\;i=1,\ldots, n-1\cr}$$
and
$$\omega=i\partial \bar\partial\left(K_1+\cdots +K_{n-1} \right)$$
which means that the K\"ahler potential for the K\"ahler metric associated to $\omega$ is $K$, where
$$K=K_1+\cdots + K_{n-1}  $$
Next,  we would like to find these $(n-1)$ potentials $K_i$.
\par
The technical material  one needs for  this task comes from the Borel--Weil construction, cf. Nash \ref{11}. 
\par
The Borel--Weil construction takes the flag manifold
$${G\over T}$$
and, if $r$ is the rank of $G$, realises the $r$ fundamental representations as spaces of holomorphic sections
of the $r$ line bundles
$$L_i,\;i=1,\ldots, r$$
over the flag manifold
$${G_{\bf c}\over B}={G\over T}=F_n.$$
These are the same line bundles $L_i$ we encountered above whose Chern classes were given by   $c(L_i)=[\omega_i/2\pi ]$.
\par
 Lie group theory tells us that each  fundamental representation---is irreducible---and is determined  by a simple root of $G$,
 and is realised on an exterior product space\foot{The holomorphic sections  mentioned above constitute   the space $H^0(F_n;L_i)$, for which  one has
  the isomorphism
$$H^0(F_n;L_i)\simeq \Lambda^i (\bc^d).$$
} 
 $$\Lambda^i (\bc^d).$$
Here $d=\dim_{\bf c} G_{\bf c}/B$.
\par
Thus we have  established  that there is  a direct  route from the K\"ahler form $\omega_i$ of $F_n$ to the 
exterior product space
 $$\Lambda^i (\bc^d),\quad d=\dim_{\bf c} F_n.$$
 Now we wish to find $K_i$.
 \par
 Suppose that we can find  a $G$ invariant  metric on the space of sections  of $L_i$; this should give a $G$ invariant metric on $ F_n$,  if we can
 pull it back to $F_n$ by some appropriate map. This pull back should allows us to identify the desired K\"ahler potential  $K_i$.   
\par
The key to finding such a $G$ invariant  metric is to embed $F_n$ into projective space via a map
$$e:F_n\longrightarrow P\left(\Lambda^i (\bc^d)\right).$$
Now take a point $p$ on the RHS side with homogeneous coordinates $[p_0:p_1:\cdots: p_{max}]$, where $max=\dim_{\bf c} \Lambda^i (\bc^d)$.  This point $p$, belonging to the fibre, being just a line in 
${\bf C}^{max}$, 
represents    a vector ${\bf V}$ possessing  a Hermitian, $G$-invariant, norm  $H_{L_i}$, where 
$$H_{L_i}({\bf V})^2=\Vert {\bf P}\Vert^2=\vert p_1\vert^2+\cdots+ \vert p_{max}\vert^2.$$
We note in passing that the $G$ invariance and the irreducibility of the Borel--Weil fundamental representation $L_i$,  
together with Schur's lemma, mean that $H_{L_i}$ is unique\foot{For let  $H^1_{L_i}$ and $H^2_{L_i}$ be two such metrics.  and ${\bf V, \, W}$
be arbitrary vectors. Then, for some invertible ${\bf A}$, one can assert that   
$$H^2_{L_i}({\bf V},{\bf W})=H^1_{L_i}({\bf V},{\bf A\kern-1pt W})$$
and, for  $g\in G$, $G$  invariance of the two metrics immediately  reveals that
$${\bf A }g=g{\bf A}.$$
Thus Schur's lemma requires that 
$${\bf A}=\lambda {\bf I}$$
for some constant $\lambda$.}  up to a multiplicative constant. 

The pull back of $H_{L_i}$ to $F_n$ is
$$e^*H _{L_i}$$
and we shall  denote it  by $h_i$, and the coordinates on $F_n$ by $z$.
Summarising, we have
$$\eqalign{ 
         e:F_n&\longrightarrow P\left(\Lambda^i (\bc^d)\right)\cr
e^* H_{L_i}&=h_i\cr
  e^*p_i&=z_i. \cr}
$$
Now for the length of a vector
$${\bf v}=\left[\matrix{v_1\cr \vdots\cr v_n\cr} \right]$$
$F_n$, up to a multiplicative constant,  we simply have  
$$h_i({\bf v})=\sqrt{\vert v_1\vert^2+\cdots+ \vert v_{max}\vert^2}=\Vert {\bf v}\Vert. $$   
Note that $h_i$ is a {\it holomorphic metric} on $L_i$ and so can be used to give  the well known  holomorphic  connection
$$
\eqalign{A_i&=h_i^{-1}\bar\partial h\cr
                &=\partial \ln h_i \cr} \quad (\hbox{$h_i$ is a 1 dimensional object})
$$
on $L_i$. Now $A_i$ has type $(0,1)$ and so its curvature $F_i$  has type $(1,1)$; one has
$$F_i=(\partial +\bar\partial)A_i=\partial\bar\partial\ln h_i. $$
But $F_i$ gives the Chern class of $L_i$ via 
$$\eqalign{c_1(L_i)&=[i F_i/2\pi]\cr
                            &=i\partial\bar\partial\ln h_i\cr}$$
But we knew already that 
$$c_1(L_i)=[i \omega_i/2\pi]$$
so, moving inside the cohomology class  we can set
$$\omega_i=\partial\bar\partial\ln h_i$$
and we identify  at once  the K\"ahler potential $K_i$: it is given by
$$K_i=\ln h_i.$$
Our last step is to focus  on  the nested subspaces 
$$V_1\subset V_2\subset\cdots \subset V_k\subset V,\hbox{ with } \dim V_i =i$$
of the flag manifold,  and then evaluate $h_i$ on a judiciously chosen vector ${\bf v}$.
\par
This nesting property is catered  for by regarding   the columns of the our unipotent lower triangular  coordinate matrix
$$n_-$$ 
as a series of vectors. For the general case of $F_n$ we have
$$n_- =\left[\matrix{
1 & 0 & 0 & \dots & 0 & 0 \cr
z_1 & 1 & 0 & \dots & 0 & 0 \cr
z_2 & z_3 & 1 & \dots & 0 & 0 \cr
\vdots & \vdots & \vdots & \ddots & \vdots & \vdots \cr
z_{d_n-n+2} & z_{d_n-n+3} & z_{d_n-n+4} & \dots & z_{d_n} & 1 \cr
}\right]_{n\times n}
\hbox{where }\left\{\eqalign{d_n&={n(n-1)\over 2}\cr
                                                                                  &=\dim_{\bf c} F_n.\cr}\right.
$$
So now let the first $(n-1)$ columns of the matrix be the vectors ${\bf v_1,  v_2, \cdots, v_{n-1}}$ so that\foot{One should look carefully at these
vectors: the suffices on the $z_i$ do not increase as one descends a column but rather as one  traverses a row---this convention is  rather messy but
I think it may be the convention  that is in use in the rest of the literature.}  
$${\bf v_1}=\left[\matrix{1\cr
                                       z_1\cr
                                       z_2\cr
                                       z_4\cr
                                       \vdots\cr
                                       z_{d_n-n+2}\cr}\right]_{n\times 1}\,
                                       {\bf v_2}=\left[\matrix{0\cr
                                                                          1\cr
                                                                          z_3\cr
                                                                          z_5\cr
                                                                          \vdots\cr
                                                                          z_{d_n-n+3}\cr} \right]_{n\times 1}\, \hbox{etc.}$$

These vectors  ${\bf v_1,  v_2, \cdots, v_{n-1}}$  can  be wedged together to provide natural definitions  for the nested  subspaces $V_i$ above.
This works as follows: let
$$V=\bc^n$$
be  a large vector space from which we  select a fixed vector   
$$v_1$$
then define   $V_1$ by
$$V_1=\{v\in V\mid v\wedge v_1=0 \}$$
this gives us a 1 dimensional space. Next we fix two vectors
$$v_1,v_2\in V$$
and define $V_2$ by
$$V_2=\{v\in V\mid v\wedge v_1\wedge v_2=0 \}$$
we note that $V_2$  is  two  dimensional
and that
$$V_1\subset V_2.$$
Continuing in this way we compile the obvious list
$$\matrix{ {\bf v_1}\hfill &\hbox{defines }V_1\cr
{\bf v_1\wedge v_2}\hfill &\hbox{defines }V_2\cr
{\bf v_1\wedge v_2\wedge v3}\hfill  &\hbox{defines }V_3\cr
\vdots                                                          &\vdots\cr
{\bf v_1\wedge v_2\wedge \cdots v_{n-1}}\hfill  &\hbox{defines }V_{n-1}\cr
\cr}
$$ and we have the desired  flag manifold nesting 
$$V_1\subset V_2\subset\cdots \subset V_{n-1}\subset V,\hbox{ with } =\dim V_i =i.$$
But
$${\bf v_1\wedge v_2\wedge \cdots v_i}\in \Lambda^i \bc^n$$
so ${\bf v_1\wedge v_2\wedge \cdots v_i}$  belongs to the   space  of holomorphic  sections  of
the line bundle $L_i$ for which we established that\foot{It may be helpful to note that the highest weight vector which determines the
irreducible representations space $\Lambda^i \bc^n$, is simply ${\bf e_1\wedge\cdots\wedge e_i}$, where the ${\bf e_i}$ are basis vectors
of $\bc^n$.}
 $$c_1(L_i)=\left[ i\omega_i/2\pi\right]$$
and also that
$$\eqalign{\omega_i&=\partial\bar\partial \ln h_i\cr
                                &=\partial \bar\partial K_i.\cr}
$$
 We need  to obtain a formula for $K_i$ in  terms of coordinates on $F_n$; but these coordinates are the entries in the matrix
 $$n_-$$
and these  are now arranged in columns, which naturally  wedge together to give the large wedges $ {\bf W_i}$, where   
$${\bf W_i}={\bf v_1\wedge v_2\wedge \cdots v_i}\in \Lambda^i \bc^n$$
and, evaluating $h_i$, gives
$$h_i\left({\bf W_i}\right)=\Vert{\bf W_i}\Vert^2.$$
This means that our formula for $K_i$ is just 
$$K_i=\ln \Vert {\bf W_i}\Vert^2$$
and the total K\"ahler potential for the metric on $F_n$ is $K$ where 
$$K=K_1+\cdots+K_{n-1}$$
Note that  we can easily unravel  the $z_i$ coordinates from ${\bf W_i}$ to give a more familiar expression,
as we shall now illustrate for our case $n=3$.

\par\vskip\baselineskip

\noindent
{\bf The K\"ahler potential for $F_3$}
\par\vskip\baselineskip\noindent
When $n=3$ we have
$$\eqalign{K&=K_1+K_2\cr
                    &=\ln\Vert {\bf W_1}\Vert^2+\ln \Vert {\bf W_2}\Vert^2\cr
                    &=\ln \Vert {\bf v_1}\Vert^2+\ln \Vert {\bf v_1\wedge v_2}\Vert^2.\cr}
                    $$
But
$$\matrix{{\bf v_1}=\left[\matrix{1\cr
                                       z_1\cr
                                       z_2\cr}
                                       \right]\cr
                    \cr
{\bf v_2}=\left[\matrix{0\cr
                                       1\cr
                                       z_3\cr}\right]\cr}\Rightarrow \left\{\eqalign{\Vert {\bf v_1}\Vert^2&= 1+ \vert z_1 \vert^2 +\vert z_2\vert^2\cr
                                                                                                        \Vert {\bf v_1\wedge v_2}\Vert^2&=1+\vert z_3\vert^2+ \vert z_1 z_3-z_2 \vert^2 \cr}\right.
$$
so, in agreement with the expression used  at the very beginning of this entry,  the full K\"ahler potential is
$$K=\ln (1+ \vert z_1 \vert^2 +\vert z_2\vert^2)+\ln (1+\vert z_3\vert^2+ \vert z_1 z_3-z_2 \vert^2). $$

\vskip 0pt 
\appendixsection{B. The  permutation length  $l(w)$}
\par
Consider the case where $G=U(n)$, so that the Weyl group is a  permutation group 
$$W\simeq S_n.$$
Then let $w\in W$ be given by the permutation
$$w=(i_1 i_2\ldots i_n)$$
where $ i_1 i_2\ldots i_n$ are the integers $1,2,\dots, n$ in some order: in other words
$$w(1)=i_1,\,w(2)=i_2, \ldots,\, w(n)=i_n.$$
The {\it inversion number } $d(w)$ of a  permutation $w$ is the number of pairs 
$$w(i),w(j)\;\hbox{ with }i<j\;\hbox{ such that }w(i)> w(j).$$
For example, consider the cases  $n=3$ and $n=5$, and the permutations
$$w_1=(231) \hbox{ and } w_2=(45321).$$
These have inversion numbers\foot{There are two  pairs in  the count for $w_1$, namely $(2,1)$  and $(3,1)$; while there are $9$ pairs for $w_2$, these being
$(4,.3),(4,2),(4,1),(5,3),(5,2),(5,1),(3,2),(3,1),(2,1)$.
}
$$d(w_1)=2\hbox{ and }d(w_2)=3+3+2+1=9.$$
\par
Finally in the Bruhat decomposition, $d(w)$ is  called the {\it length} of $w\in W$ and is written $l(w)$.

\vfill\eject

\startreferences
\refentry
 Cabibbo N.,  Unitary symmetry and leptonic decays,  Phys. Rev. Lett.,  {\bf 10}, 531--533,  (1963).
 \refentry
Kobayashi M. and Maskawa T., 
  CP violation in the renormalisable theory of weak interactions,  Prog. Theor. Phys.,   {\bf 49}, 652--657, (1973). 
\refentry
Alkofer R., {\it et al}.
Quark masses and mixings in minimally parameterized
UV completions of the Standard Model,
{\it Annals of Physics}, {\bf 421}  (2020),  168282,
arXiv: 2003.08401 [hep-ph].
\refentry
Dolan B. P.,  Fixed points of the CKM matrix
renormalization group running to all
orders in perturbation theory, [arXiv:2608.28149 [hep-ph]].
\refentry
Dolan B. P.,   Fixed points of the renormalisation group running of quark and fermion mixing  matrices in the Standard Model  and beyond, {\it Phys. Rev. D}
{\bf 113} (2026), 095043, [arXiv:2601.02452 [hep-ph]].
\refentry 
Buchstaber V. M. and Terzi\'c S., 
The foundation of $(2n,k)$ manifolds,  
Sbornik: Mathematics,  {\bf 210}, 508-549, (2019). 
\refentry
Atiyah M. F.,  Convexity and commuting Hamiltonians,  Bull. Lond. Math. Soc.,  {\bf 16}, 1--15, (1982). 
\refentry
Atiyah M. F.,   Angular momentum, convex polyhedra and algebraic geometry,  Proc. Edin. Math. Soc., {\bf 26}, 121--38, (1983). 
\refentry
Kirwan F. C.,  Convexity properties of the moment mapping, III,  Invent. Math., {\bf 77},  547--552,  (1984). 
\refentry
Hatcher A., {\it Algebraic topology},  Cambridge University Press,  (2002).
\refentry
Nash C., Differential Topology and Quantum Field Theory, Academic Press, (1991).

\vskip -5pt

\bye